# Negligible thermal contributions to the spin pumping signal in ferromagnetic metal-Platinum bilayers

Paul Noël[1], Maxen Cosset-Cheneau[1], Victor Haspot[2], Vincent Maurel[3], Christian Lombard[3], Manuel Bibes[2], Agnès Barthelemy[2], Laurent Vila[1], Jean-Philippe Attané[1]

[1]Univ. Grenoble Alpes, CNRS, CEA, Grenoble INP, IRIG-SPINTEC, F-38000 Grenoble, France

[2]Unité Mixte de Physique, CNRS, Thales, Université Paris-Sud, Université Paris-Saclay, Palaiseau, France

[3]Univ. Grenoble Alpes, CEA, CNRS, IRIG-SYMMES, F-38000 Grenoble, France

**Abstract:** Spin pumping by ferromagnetic resonance is one of the most common technique to determine spin hall angles, Edelstein lengths or spin diffusion lengths of a large variety of materials. In recent years, rising concerns have appeared regarding the interpretation of these experiments, underlining that the signal could arise purely from thermoelectric effects, rather than from coherent spin pumping. Here, we propose a method to evaluate the presence or absence of thermal effects in spin pumping signals, by combining bolometry and spin pumping by ferromagnetic resonance measurements, and comparing their timescale. Using a cavity to perform the experiments on Pt\Permalloy and $La_{0.7}Sr_{0.3}MnO_3$\Pt samples, we conclude on the absence of any measurable thermoelectric contribution such as the spin Seebeck and anomalous Nernst effects at resonance.

## I Introduction

Spinorbitronics is based on the interconversion of charge currents into spin currents, by the spin Hall Effect in bulk materials or by the Edelstein effect at surfaces and interfaces. Determining the spin to charge current conversion efficiency, *i.e.*, the spin Hall Angle or the inverse Edelstein length, and the spin diffusion length, is a key point to understand experimental results and develop applications. Materials with large spin Hall angles are indeed required for a variety of foreseen applications such as 3-terminal SOT MRAM [1], Magnetoelectric Spinorbit Logic [2] or Terahertz emitters [3]. Since the early 2000's several techniques have emerged to evaluate this conversion efficiency, using lateral spin valves [4], Spin Seebeck measurements [5], spin pumping FMR (SP-FMR) [6], magneto-optical Kerr effect measurements [7], etc. The spin pumping FMR technique has been widely used to evaluate the spin Hall angles and inverse Edelstein lengths of a large numbers of materials, including heavy metals [6,8,9], semiconductors [10], Rashba interfaces [11, 12] and topological insulators [13,14,15,16]. The reason of such a wide use relies on the compatibility of SP-FMR with any multilayer stack containing a ferromagnet, and the absence of any complex and costly nanofabrication process. Moreover ferromagnetic resonance was an already widely used spectroscopy technique to evaluate the dynamical and non-dynamical properties of ferromagnetic thin films

Even for a widely studied material as Platinum the estimated values of spin diffusion length and spin Hall angle determined by different techniques spread over more than one order of magnitude, with spin diffusion length ranging from 1.2 nm [17] to 11 nm [18] and spin Hall angle from 1.2% [19] to 38.7% [20]. This large discrepancy can be partially explained by differences in Pt resistivity or accounted for by interface-related phenomena such as the spin memory-loss [9,21,22,23,24], but it remains mostly unexplained [25]. In this particular context, concerns regarding the reliability of the SP-FMR technique have been pointed out. A thermal gradient could indeed arise at the ferromagnetic resonance, due to the energy absorption in the ferromagnetic layer [26,27,28,29] and give rise to several thermoelectric and spin-caloritronics contributions – in particular the anomalous Nernst effect (ANE) and the spin Seebeck effect (SSE) – that would add up to the spin pumping inverse spin Hall effect (ISHE) signal. In this picture, the signal would thus be due to a combination of the ISHE signal, the spin rectification effects (SRE) and thermal effects [30].

While the separation of ISHE from SRE has already been vastly discussed and can be achieved from the angular dependence in different measurement geometries [30,31,32,33], disentangling the ISHE signal from thermal effects remains an open question. While the ordinary Seebeck effect (OSE) and ordinary Nernst effect (ONE) contributions may be extracted from angular dependences, it is not the case of the ANE and SSE contributions, which possess an angular dependence similar to that of the ISHE signal. Recent analysis of spin-pumping FMR results have even been based on the hypothesis that the observed signals are dominated by the SSE [27]. In that case the nature of the spin pumping signal would be incoherent –thermal- and the extensively used coherent spin pumping model would strongly misestimate the injected spin current [34]. It is therefore possible that due to a large contribution of incoherent spin pumping, the estimation of the Spin Hall Angle by spin pumping FMR in Pt is inaccurate. More generally, a contribution of ordinary thermoelectric effects such as the ordinary Nernst effect is also a possible explanation to the discrepancy between the lack of spin charge conversion in Silver Bismuth bilayer measured by means of Longitudinal Spin Seebeck Effect (LSSE) [35], and the large one measured by means of spin pumping [11,12]. Therefore, estimating possible thermoelectric or spin-caloritronics contributions is of high importance to obtain an accurate evaluation of the spin charge conversion efficiency.

## II Experimental procedure

Figure 1 a) depicts the dynamical spin injection process as described by Tserkovnyak *et al.* [36], which is the model used to analyze the signal observed in spin pumping FMR measurements. As suggested by Yamanoi *et al.* [27], the additional dissipation at the FMR could lead to the appearance of a voltage along the x direction. The absorption at resonance would lead to a temperature increase of the ferromagnet, and thus to a thermal gradient perpendicular to the layers. This thermal gradient would lead to the injection of a pure spin current along Z towards the non-magnetic material, converted by ISHE into an electric field along X through a process known as the longitudinal Spin Seebeck effect as seen in figure 1.b. Owing to the existence of a thermal gradient, the anomalous Nernst effect in the FM layer could also appear, creating an electric field along X as depicted in figure 1c.

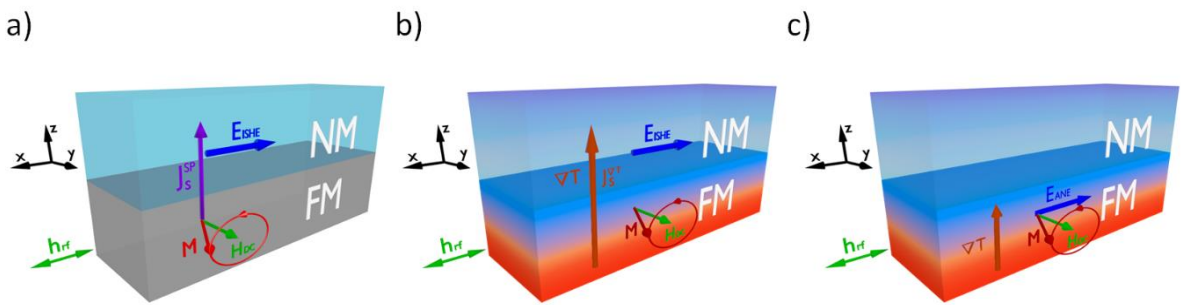

*Figure 1: Schematic representation of the possible spin injection mechanisms at the FMR and of thermal gradient related effects. a) Dynamical spin injection. Because of the magnetization precession, the spin current $J_S^{SP}$ is injected from the FM layer towards the NM layer. An electromotive force $E_{ISHE}$ then arises along X, due to ISHE, which can be detected as a voltage in open circuit. b) Spin injection due to the thermal gradient. At the ferromagnetic resonance the temperature of the FM layer increases creating a thermal gradient $\nabla T$ along the z direction and thus a thermal spin current injection $J_S^{\nabla T}$. This spin current is then converted into an electromotive force $E_{ISHE}$. c) Thermal gradient within the ferromagnet could give rise to an anomalous Nernst related electromotive force $E_{ANE}$.*

We propose to test these hypothesis in two multilayers. The first one is a Pt\Permalloy bilayer, archetypal of spin pumping ISHE experiments [6, 8, 17, 19], with a large ANE coefficient in Permalloy (Py) [37], and the second one is a La$_{0.7}$Sr$_{0.3}$MnO$_3$\Pt bilayer, in which a large SSE contribution is expected [38]. The characteristic timescale of the FMR spin injection mechanism is the FMR precession period, which is of the order of the nanosecond. But the temperature increase timescale, the time needed to reach a thermal equilibrium, is of several seconds [39,40]: thermal and non-thermal effects have different dynamics. We thus propose a technique that can be adapted to any SP-FMR experiment to disentangle the two mechanisms, by measuring the time dependence of the spin pumping signal and of the temperature increase.

### III Results and Discussion

We performed SP-FMR measurements on a SiO$_2$\\Pt(10)\Py(20) multilayer (the numbers in parenthesis represent the thickness in nanometers), on a 2.4×0.4 mm² structure positioned in a Bruker MS5 loop gap cavity. We performed a FMR measurement at different sweeping rates, at a power of 100 mW. The scheme of the measurement is shown in figure 2 a), and consists in the measurement of the voltage at the ferromagnetic resonance. The DC magnetic field can be applied in the plane of the sample, and perpendicularly to the electrical contacts. This configuration is the parallel configuration. The sample is then turned by 180°, in a position corresponding to the antiparallel configuration. As seen in figure 2 b) and c) in both the parallel and antiparallel configurations, the signal is symmetric and independent of the sweeping time. In both cases we subtracted the offset signal and divided by the square of the radiofrequency magnetic field h$_{rf}$ to have comparable results. The amplitude of the radiofrequency field was determined by measuring the Q factor with the sample placed inside the cavity using the following equation: h$_{rf}^2$ =4PQ/500, with P the microwave power in Watt [33].

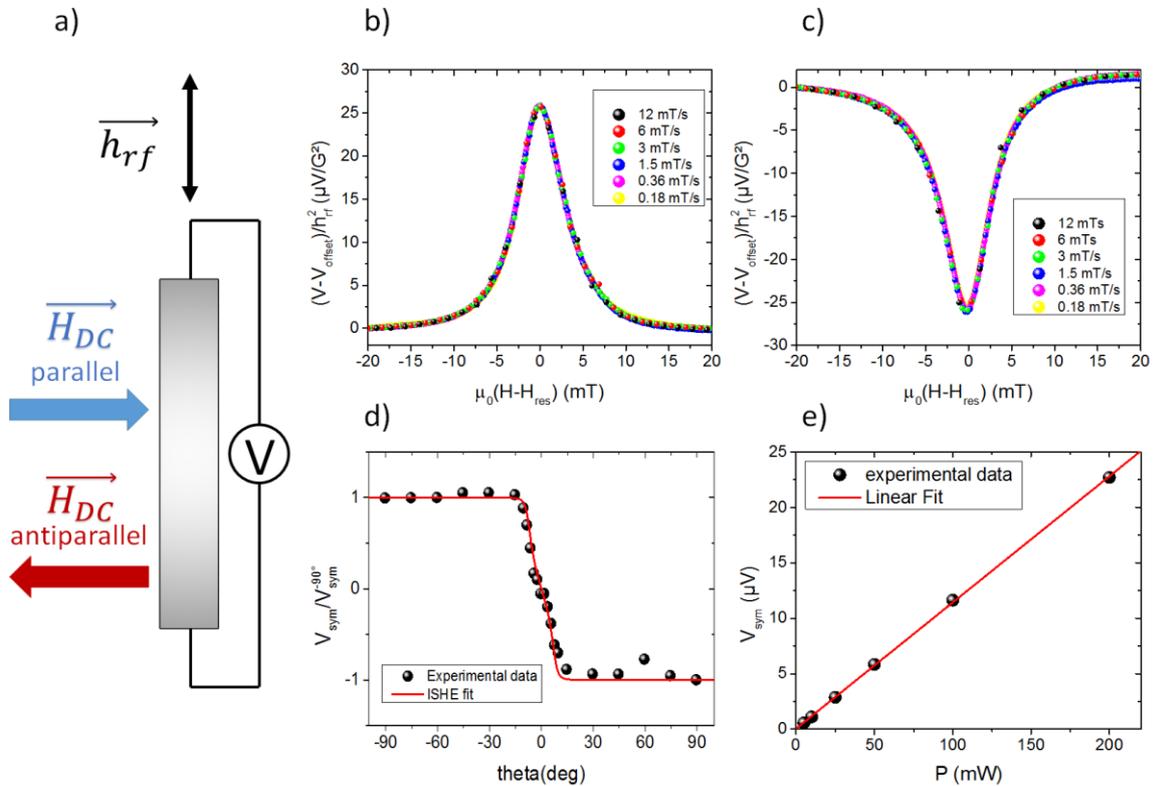

Figure 2: a) Schematic representation of the measurement device used to detect spin pumping ISHE at resonance in the Pt(10)\Py(20) sample (top view). b) Signal obtained in the in-plane parallel configuration for *various sweeping rates, normalized by the square of the rf magnetic field $h_{rf}^2$ for a power of 100mW. The Q factor was 690 and the symmetric voltage of 14.7 µV before normalization. c) Similar measurement in the in-plane antiparallel configuration. The Q factor was 537 and the symmetric voltage of -11.8µV before normalization d) Out-of-plane angular dependence of the spin signal fitted using the ISHE angular dependence provided in ref [33] e) Power dependence of the symmetric part of the signal as a function of the microwave power in the parallel configuration, the Q factor was 510.*

Regarding the possible contribution of the spin rectifications effects in Py, the out-of-plane angular dependence has also been performed (cf. figure 2 d). The obtained symmetric signal can be fitted with the ISHE angular dependence model described in reference [33] as $\sin(\Theta_M)$ with $\Theta_M$ being the magnetization angle with respect to the out of plane direction. This also excludes any contribution of the ordinary Seebeck effect, which would be field independent, and of the Nernst effect, which would depend on the applied field perpendicular to the thermal gradient. The signal is also linear with the power (cf. figure 2 e) indicating a negligible change of magnetization when increasing Power. The signal possesses the ISHE angular dependence, and there is no trace of thermal drift, which implies that if there is a thermal component to the signal, a steady state with the thermal gradient has to be reached in a characteristic time well below one second.

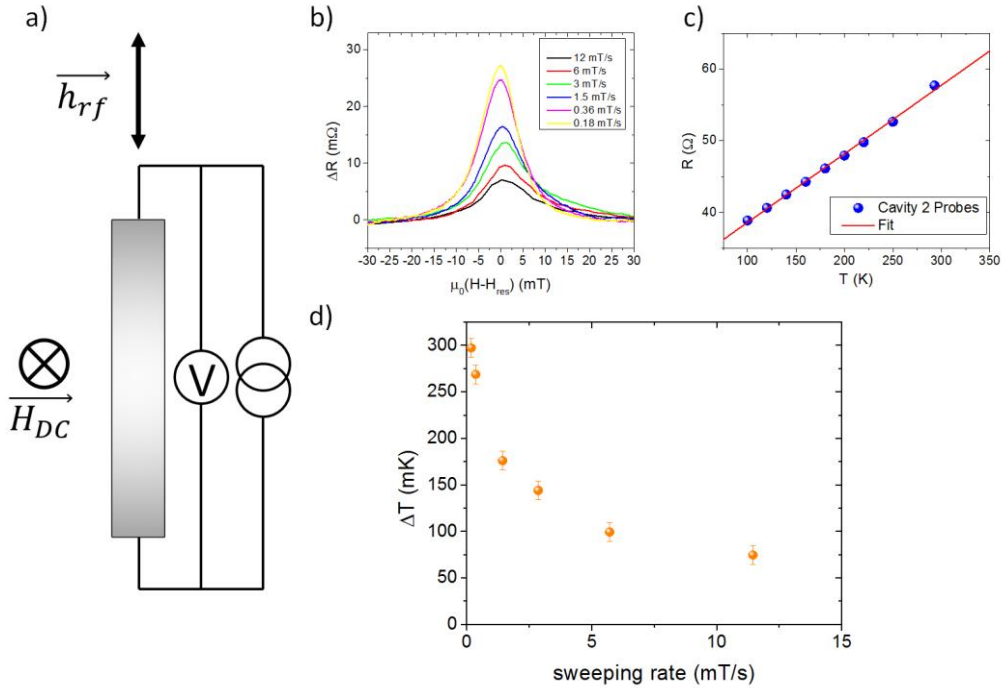

*Figure 3: a) Schematic representation of the measurement device used to detect a resistance change at resonance in the Pt(10)\Py(20) sample (top view). b) Change of the two probes resistance around the resonance field, for various field sweeping rates (with base resistance subtracted). c) Resistance of the sample as a function of the temperature, measured in a 2 probes configuration on the same sample. The slope value is 96±2mΩ/K d) Temperature change as a function of the sweeping rate, estimated from the increase of resistance at resonance.*

Let us now evaluate this characteristic time. A temperature increase can occur at the ferromagnetic resonance, due to the increased microwave absorption at resonance [26,39,40,41]. To evaluate the time dependence of this effect, we adopt the measurement scheme shown in figure 3 a), where the field is applied out of plane to avoid ISHE or SRE voltage contribution. We used a fixed DC current of 1 mA, fixed power of 100 mW and measured the change of resistance at resonance, known as the bolometric effect. As can be seen on figure 3 b) we observe an increase of resistance at resonance;

more importantly, this increase is highly dependent on the field sweeping rate. The resistance increases from 7.2±1 mΩ for a sweeping rate of 12 mT/s, to 28.5±1 mΩ for a sweeping rate of 0.18 mT/s. The result is in stark contrast with fig. 2 where the signal is independent on the sweeping rate. The temperature increase characteristic time is thus of several seconds, as the time spent near resonance at a sweeping rate of 0.18 mT/s is of 20s for a linewidth of 3.5 mT. This timescale is similar to what has been observed in previous Electrically Detected FMR experiments [39,40]. A similar result was obtained with in plane field.

We estimate the corresponding temperature increase using the temperature dependence of the resistance of the sample measured inside the cavity (cf. figure 3c). The linear behavior of the resistance leads to a temperature dependence of 96±2 mΩ/K in a range from 100K to Room temperature. The temperature increase as a function of the sweeping rate is shown in figure 3d. The maximum temperature increase is of 297±10 mK for the slowest sweeping rate, and of 75±10 mK for the fastest. The temperature increase is thus found to be strongly dependent on the sweeping time, the temperature stabilization is not reached even after several seconds near resonance. Therefore, any effect originating from a temperature change should vary with the sweeping time. The spin signals measured in the configuration of fig. 2 being totally independent of the sweeping time, we can conclude that in $SiO_2\backslash\backslash Pt(10)\backslash Py(20)$ the longitudinal spin Seebeck effect and anomalous Nernst effect are negligible, and that the observed signals are due to coherent spin pumping. This comparative measurement can also be performed using a coplanar waveguide setup where bolometric effects with a time constant of several seconds were previously observed [40]. Some hundreds of mK might appear to be a small temperature increase when compared to the several Kelvins usually used in LSSE measurements. Nonetheless due to the large thickness of the substrate compared to the Ferromagnetic thin films, the temperature difference between the bottom and the top of the ferromagnetic layer is usually below 1mK in LSSE measurements [42]. For the ordinary Nernst contribution a temperature difference of only a few mK could also give rise to signals of some µV [43].

In order to verify this lack of thermal contribution in Pt we also performed a combined bolometric and SP-FMR measurements on a $LSAT\backslash\backslash La_{0.7}Sr_{0.3}MnO_3(13.8)\backslash Pt(8.2)$ sample, measured along the [100] direction. $La_{0.7}Sr_{0.3}MnO_3$ (LSMO) possesses a high resistivity compared to Permalloy and Platinum moreover the LSMO\Pt structure is expected to possess a smaller ANE coefficient but a larger SSE contribution than Py\Pt as demonstrated in Longitudinal Spin Seebeck experiments [38]. Therefore the possible contribution of SSE in this multilayer is expected to be enhanced compared to Pt\Py. In figures 4b and 4c similarly to the case of Pt\Py we can see that the thermal equilibrium is still not reached even for the slowest sweeping rate, the total temperature increase is of comparable amplitude and up to 199±3mK. As can be seen in figure 4e and 4f, the obtained spin signal is independent on the sweeping rate. Here again, this shows that in this system the ANE and SSE contributions are negligible compared to the spin pumping ISHE signal. As already observed in LSMO layer, in the out of plane configuration the resonance peak is slightly asymmetric, because of magnetic homogeneities [44], this leads to an asymmetry of the bolometric response as observed in figure 4 b). This is not the case in Permalloy, because the small grain size leads to a strong exchange narrowing [45] and thus to a small inhomogeneity. This inhomogeneity is not observed in plane and ISHE signal in figure 4 e) and f) are consequently symmetric.

The NM and FM stacking order is inverted in the Pt\Py sample, which is why the spin signal is of opposite sign when compared to LSMO\Pt and to previous results on Co\Pt [9], as expected for ISHE symmetries [46]. The normalized ISHE signal is the ISHE voltage divided by the square of the rf field, the width and the total resistance of the device. The obtained values are of 0.78 mV.G$^{-2}$Ω$^{-1}$m$^{-1}$ in LSMO\Pt and 1.11 mV.G$^{-2}$Ω$^{-1}$m$^{-1}$ in Pt\Py, similar to the value of 0.85 mV.G$^{-2}$Ω$^{-1}$m$^{-1}$ to 1.13 mV.G$^{-2}$Ω$^{-1}$m$^{-1}$ that was previously reported in SiO$_2$\\Co\Pt of similar thicknesses at X-band [9] indicating a similar injected spin currents in these three structures.

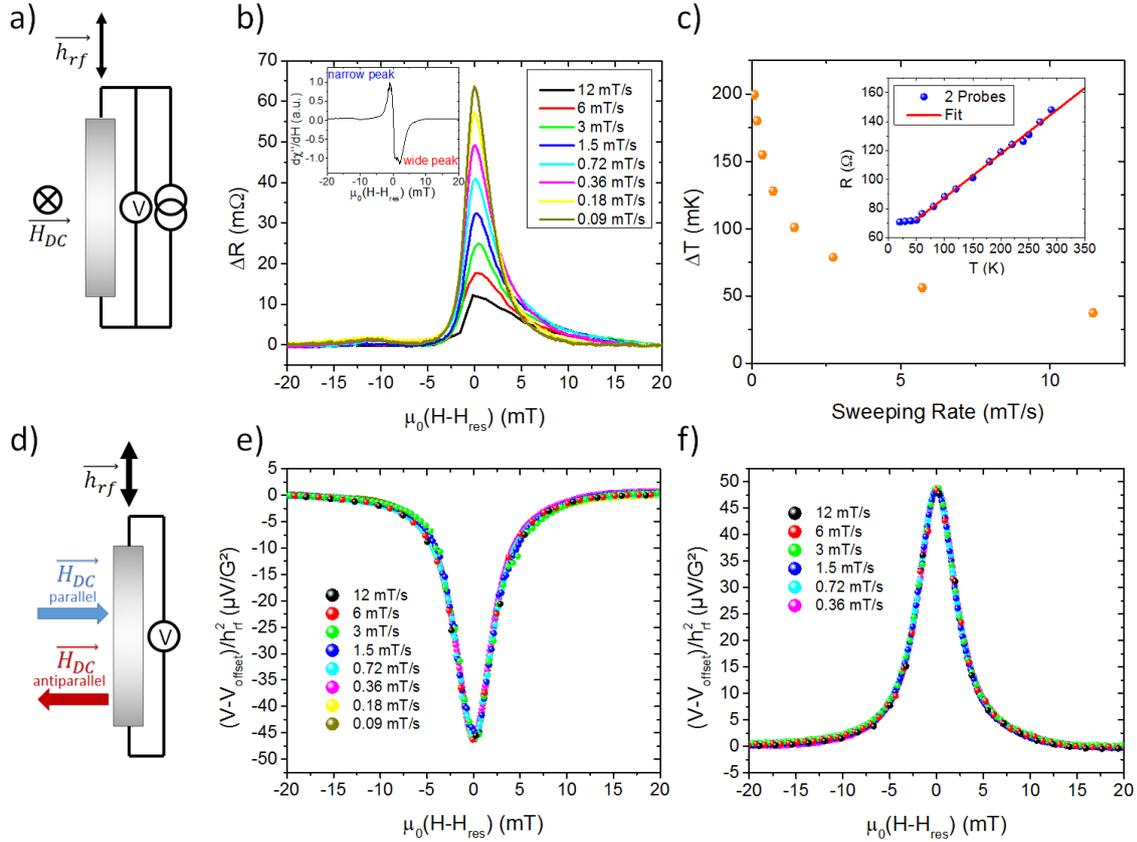

*Figure 4: a) Schematic representation of the measurement device used to detect resistance change in the LSAT \\La$_{0.3}$Sr$_{0.7}$MnO$_3$(13.8)\Pt(8.2) sample (top view). b) Change of resistance for various sweeping rate. Inset shows the FMR response out of plane in LSMO is asymmetric, with a narrow and a wide peak c) Temperature change as a function of sweeping rate. Inset shows the resistance as a function of temperature, slope is of 303±6 mΩ/K. d) Schematic representation of the measurement device used to detect spin pumping (top view). e) Signal obtained in the in-plane parallel configuration for various sweeping rates, normalized by the squared rf magnetic field $h_{rf}^2$. The Q factor was 510 and the symmetric voltage of -18.8 µV before normalization. f) Similar measurement in the in-plane antiparallel configuration. The Q factor was 269 and the symmetric voltage of 11.8 µV before normalization.*

Another experiment can be done to demonstrate the absence of thermal contribution to the spin pumping signal. The sample was placed in the parallel configuration and the external field was swept as fast as possible from 20 mT below the resonance to the resonance field H$_{res}$ at a fixed rf power of 100 mW. The sweeping rate in this experiments was limited to 1 mT/s to avoid a large overshoot of the field when stopping at the resonance field and thus allowing a fast stabilization of the field comparable to our time resolution.

In a first step, a 5 mA current is applied in the sample, so that the signal variations correspond mostly to resistance variations. The voltage resulting due to Ohm's Law is of 5 µV/mΩ using a current of 5mA while the total spin pumping signal is of around 10 µV at a power of 100 mW. The results, shown in figure 5, exhibits a resistance increase when reaching the resonance field. The time constant of the temperature increase is of around 10 seconds. In a second step, the same experiment is performed in the open circuit conditions commonly used for spin pumping experiments. In that case, the maximal signal is obtained immediately after reaching the resonance field. This implies that the signal measured in open circuit conditions is not linked to the slow temperature increase at resonance but to the fast dynamical spin injection mechanism.

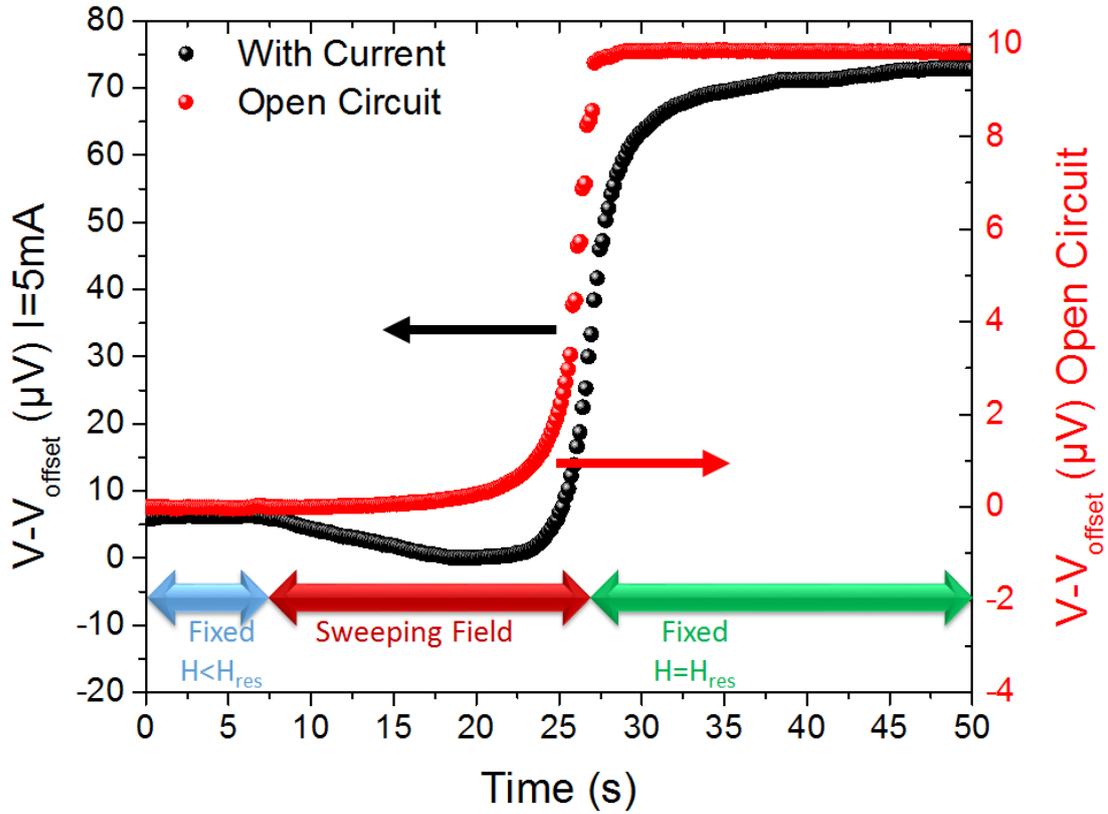

*Figure 5: Change in the measured output voltage as a function of time during the sweeping of the field from out of resonance to resonance in the SiO2\\Pt(10)\Py(20) sample for parallel to the plane configuration with an input current of 5 mA (in black) and in open circuit conditions (in red).*

In this study we focused on the possible thermal effects contributions at resonance. As previously shown there is also in our measurements an offset signal (existing at and out of resonance), that is usually subtracted, and which can also be of thermal origin [29]. The change of temperature due to microwave absorption out of resonance is orders of magnitude larger than that due to the resonance [40], and it can lead to larger thermal gradients. In order to study that point we used a technique very similar to that shown in figure 5, but out of resonance.We thus compared the evolutions of the resistance of the sample and of the offset voltage when increasing the power entering the cavity. We observed a temperature change more than one order of magnitude larger than the change observed at resonance at a similar power, and a similar evolution for both the resistance and offset voltage in both samples. These observations confirm that the offset signal has a thermal origin, and that actually the temperature increase of the sample is mostly due to microwave absorption, independently of the resonance. Note that this signal contribution, which exists both at and out of resonance, does not contribute to the spin signal amplitude.

**IV Conclusion**

We observed in SP- FMR experiments in cavity that the temperature increase at resonance is limited to a few hundreds of mK, even at a large rf power of 100 mW, and is further reduced to dozens of mK for faster field sweeping around the resonance. Moreover, regarding the angular dependencies and the absence of link between the detected signal and the temperature increase at resonance, we can conclude that the SSE and ANE are absent in the signals at resonance for both SiO$_2$\\Pt\Py and LSAT\\LSMO\Pt multilayers, and that only dynamical spin injection is involved. The experiment presented here can be generalized to any system measured using spin pumping, and in particular to Rashba interfaces and topological insulators. Indeed, these two groups of materials gather a large number of materials of interest for spinorbitronics but possess a very high thermoelectric figure of merit [47]. For example in Bi, Bi$_2$Se$_3$, or Bi$_2$Te$_3$ it could give rise to non-negligible thermal signals, unrelated to the spin-charge interconversion. It might also be a useful way to further study some recently evidenced thermal effects such as the valley Nernst effect [48] or to acquire a better understanding of the nature of the spin pumping signal in antiferromagnets [49].

**Acknowledgments:**
We acknowledge the financial support by ANR French National Research Agency Toprise (ANR-16-CE24-0017), ANR French National Research Agency OISO (ANR-17-CE24-0026) the Laboratoire d'excellence LANEF (ANR-10-LABX-51-01) and European Commission via the TOCHA project H2020-FETPROACT-01-2018 under Grant Agreement 824140. We are grateful to the EPR facilities available at the national TGE RPE facilities (IR 3443). We thank Jean-François Jacquot and Serge Gambarelli for their help and advices on the FMR measurement setup.

Data availability:
The data that support the findings of this study are available from the corresponding author upon reasonable request.